\documentclass{iaus}
\usepackage{graphicx,natbib}

\title[Tidal Debris posing as Dark Galaxies] 
{Tidal Debris posing as Dark Galaxies}

\author[Duc et al.]   
{Pierre--Alain Duc$^1$,
 Fr\'ed\'eric Bournaud$^1$ \break \and Elias Brinks$^2$}

\affiliation{$^1$Laboratoire AIM, DSM/CEA -- CNRS -- Universit\'e Paris Diderot,
DAPNIA/SAp, CEA--Saclay, 91191 Gif sur Yvette cedex, France \break email: paduc@cea.fr\\[\affilskip]
$^2$Centre for Astrophysics Research, University of Hertfordshire, College Lane, \break Hatfield AL10 9AB, UK \break email: E.Brinks@herts.ac.uk}

\pubyear{2007}
\volume{244}  
\pagerange{}
\date{?? and in revised form ??}
\jname{Dark Galaxies and Lost Baryons}
\editors{}
\begin{document}

\maketitle

\begin{abstract}
Debris sent into the intergalactic medium during tidal collisions has received  much attention as it can tell us about several fundamental properties of galaxies, in particular their missing mass, both in the form of cosmological Dark Matter and so-called Lost Baryons.

High velocity encounters, which are common in clusters of galaxies, are able to produce faint tidal debris that may appear as star--less, free floating HI clouds. These may be mistaken for Dark Galaxies, a putative class of gaseous, dark matter (DM) dominated, objects  which for some reason never managed to form stars. VirgoHI21, in the Virgo Cluster, is by far the most spectacular and most discussed  Dark Galaxy candidate so far detected in HI surveys. We show here that it is most likely made out of material expelled 750~Myr ago from the nearby spiral galaxy NGC 4254 during its fly--by at about 1000~km~s$^{-1}$ by a massive intruder. Our numerical model of the collision is able to reproduce the main characteristics of the system: in particular  the absence of stars, and  its prominent velocity gradient.  Originally attributed to the gas being in rotation within a massive dark matter halo, we find it instead  to be consistent with a combination of simple streaming motion plus projection effects (Duc \& Bournaud, 2007). 

Based on our multi-wavelength and numerical studies of galaxy collisions, we discuss several ways to identify a tidal origin in a Dark Galaxy candidate such as optical and millimetre--wave observations to reveal  a high metallicity and CO lines, and more importantly,  kinematics  indicating  the absence of a prominent Dark Matter halo. We illustrate the method using another HI system in Virgo, VCC~2062, which is most likely  a Tidal Dwarf Galaxy (Duc et al., 2007). 

Now, whereas tidal debris should not contain any dark matter from the halo of their parent galaxies, it may exhibit missing mass in the form of dark baryons, unaccounted for by classical observations, as recently found in the collisional ring of NGC~5291 (Bournaud et al., 2007) and probably in the TDG VCC~2062. These ``Lost Baryons" must originally have been located in the disks of their parent galaxies.

\keywords{Galaxies:interactions , galaxies:peculiar, galaxies: dwarf , galaxies: clusters: individual: Virgo, galaxies: individual (VirgoHI21, NGC~4254, NGC~4694, VCC~2062),  (cosmology:) dark matter}
\end{abstract}

\firstsection 
\section{Introduction}
Based on multi--wavelength studies and {\em ad hoc} numerical simulations  of three specific systems, we discuss  what debris of collisions may tell about the dark side of galaxies. 
\begin{itemize}
\item  The case of VirgoHI21 illustrates the fact that  faint  tidal debris expelled from galaxies may be mistaken for Dark Galaxies  --- a putative class of gas--rich, but star--less, galaxies. It more generally emphasizes the thus far neglected role of high--velocity encounters  (Sec.~\ref{sec:debris}). 
\item The case of VCC~2062, also in the Virgo Cluster, provides an example of a Tidal Dwarf Galaxy that has formed out of more prominent tidal material expelled during a merger. Being massive and nearby enough, it allows a detailed study, in particular of its  kinematics, revealing  the main characteristics of tidal debris. This may be used to disentangle genuine Dark Galaxies, if they exist, from collisional debris (Sec.~\ref{sec:tidal}).
\item NGC~5291 is another case of a high--speed collision which, because it was direct,  has had dramatic effects on its gas disk. The study of the recycled objects formed in its huge HI collisional ring disclosed the unexpected presence of missing matter, probably in the form of dark baryons  (Sec.~\ref{sec:dark}).
\end{itemize}

\section{Tidal debris as fake dark galaxies}\label{sec:debris}
With the availability of unprecedented, deep blind HI surveys, a population of apparently free--floating HI clouds without any detected stellar counterpart has become apparent \citep{Davies04,deBlok05, Giovanelli07a,Kent07}. Such properties have made them candidates for being Dark Galaxies. However, previous  surveys \citep[e.g.,][]{Meyer04} had shown that these clouds often lie close to massive spirals and had most likely been expelled from their disk by dynamical processes, such as the ram pressure exerted by the intracluster-medium or tidal forces due to galaxy-galaxy interactions. In particular this first mechanism is efficient, in the environment of clusters, to strip gas without stars, contrary, as is usually believed, to tidal interactions. Collisions between galaxies which result in a merger event form long gaseous as well as stellar tidal tails that are usually co--spatial. However, depending on the initial distribution of the gas disk, the HI tail may be more extended than the optical one \citep{Hibbard94} or even slightly offset to it \citep{Mihos01}. 
Being Dark Matter dominated, the kinematics of Dark Galaxies should be characterized by pronounced velocity gradients indicative of rotation. \cite{Bournaud04} and \cite{Bekki05}  noted, however, that velocity gradients may also be present in tidal debris due to streaming motions along the tail. When observed along a particular line of sight, the sign of the gradient may even appear to reverse, as if the object were kinematically decoupled from the rest of the tail and rotating, whereas in reality this signature is due to projection effects.  We present here a set of numerical simulations addressing  in more detail the issues governing the stellar content of tidal tails and their internal kinematics, both of which have been put forward  to exclude the tidal hypothesis for the origin of Dark Galaxies.

\subsection{The role of high--speed collisions}\label{sec:collisions}
Because of the high velocity dispersion of galaxies in clusters, collisions take place in this environment at high speed and most often  do not result in mergers. Their role on the evolution of galaxies had so far been neglected as they are believed to cause little damage, unless interactions happen repeatedly, forming part of a pattern of galaxy harassment. However, as shown later, they are able to produce faint gaseous tidal tails that remained unnoticed in previous generations of HI observations but have since become visible in current deep surveys. 

\begin{figure}
\includegraphics[width=\textwidth]{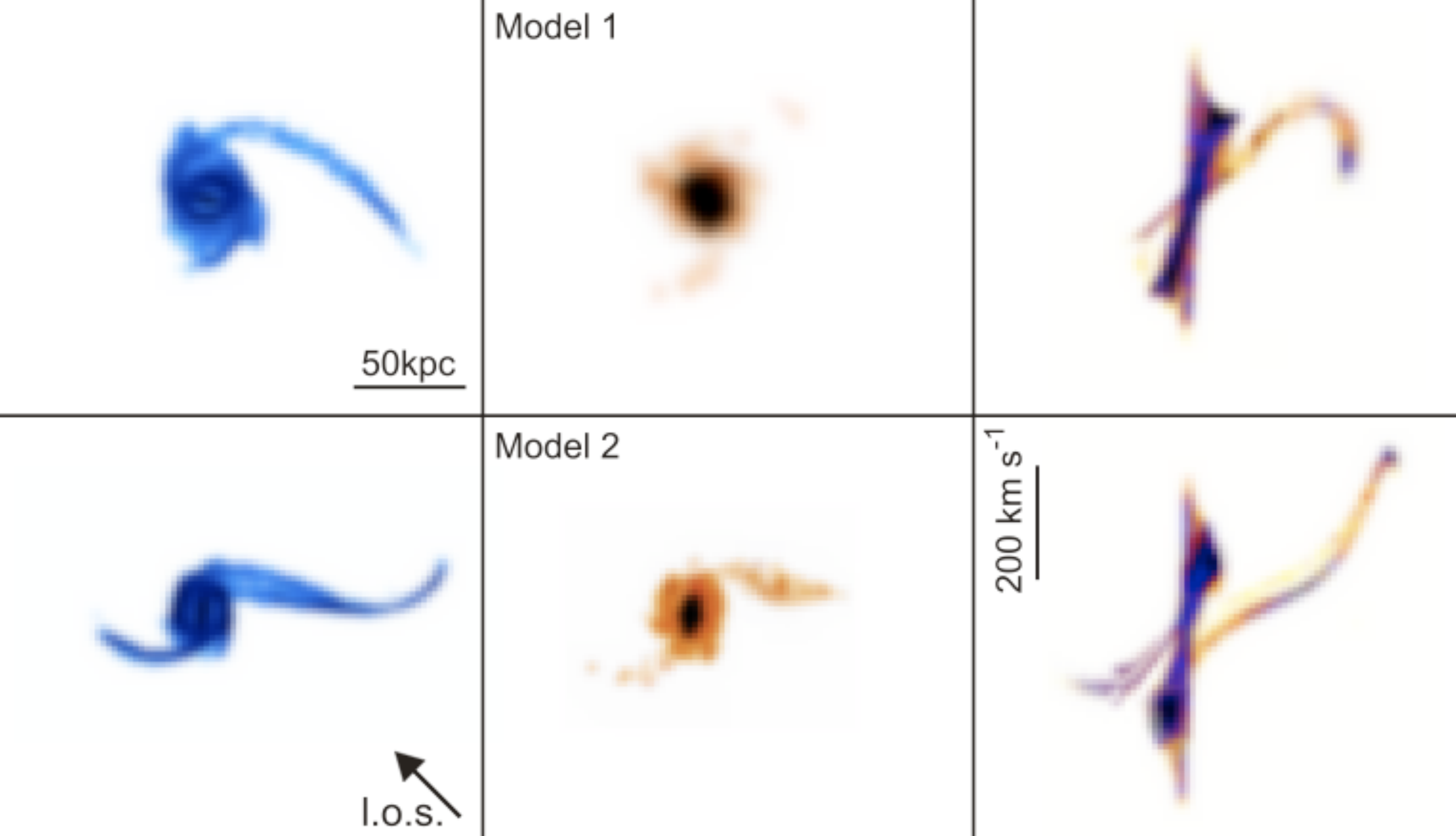}
\caption{Comparison between the effects of a high and low velocity collision on the formation of tidal tails. The upper figures show the results of a numerical simulation 300~Myr after  an encounter at a velocity at pericenter of 1050~km~s$^{-1}$. The gaseous component is shown to the left, the stellar one in the middle, and the position--velocity diagram along the tail to the right. The lower figures present the case of an encounter at a velocity of 320~km~s$^{-1}$ \citep[adapted from][]{Duc07a}. \label{fig:comp}}
\end{figure}

We have carried out a number of numerical simulations using a particle-mesh sticky-particle code \citep[see details in][]{Duc07a} in order to study the impact of the initial encounter velocity on the shape, composition and kinematics of tidal tails. Stars, gas, and dark matter haloes were modeled with one million particles each.  Looking at Figure~\ref{fig:comp} which displays the results of two different runs, one may immediately note that:

\begin{itemize}
\item High-velocity encounters with $V_\mathrm{P} \sim 1000$~km~s$^{-1}$ or more can form gaseous  tidal tails with a length comparable to those formed in low--velocity encounters,  but with a lower mass.
\item The counter--tail is fainter and shorter, hence falling back more rapidly on to the parent spiral disk. In a few hundred Myr, the system will appear as single--tailed. 
\item The stellar mass fraction in tidal debris is much lower for high velocity encounters. Most stars, if not all, remain in the parent disk, which is only weakly  disturbed. 
\item The total velocity excursion of a tidal tail, $\Delta V$, is not correlated with the encounter velocity  but is rather of the order of the parent disk circular velocity. Thus velocity spreads along tidal tails of only 200~km~s$^{-1}$ can be caused by collisions at speeds higher than 1000~km~s$^{-1}$, contrary to what intuition may tell \citep{Minchin07}.
\item The tails  exhibit apparent changes of sign in their projected velocity gradient, as clearly shown in position--velocity (PV) diagrams.
\end{itemize}

The simulated tidal tails do not necessarily have a uniform profile. Gaseous condensations may form within them without reaching the critical threshold above which star--formation occurs (rendering them visible in the optical). In moderately deep HI surveys such tidal debris will then appear as free--floating HI clouds, located at more than 100~kpc from the parent galaxies which themselves will look unperturbed; i.e., they would appear as promising Dark Galaxy candidates. We illustrate these results with the famous case of VirgoHI21 which has received much attention during this symposium.

\subsection{A numerical model of VirgoHI21}\label{sec:VirgoHI21}

\begin{figure}
\includegraphics[width=11cm]{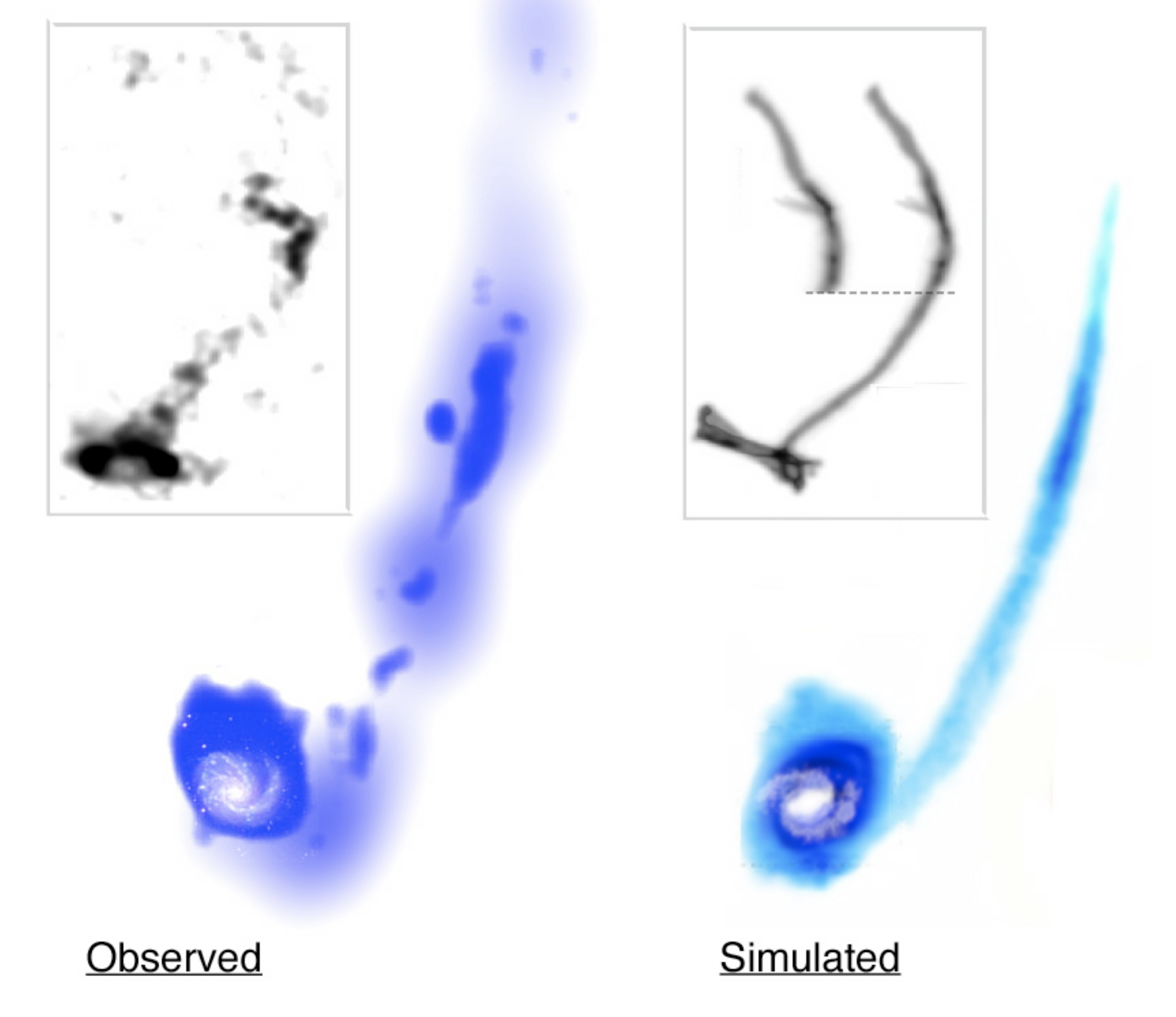}
\caption{The Dark Galaxy candidate VirgoHI21. {\it Left:} The observed system as seen in HI, a combination of the data set obtained at the Westerbork SRT by \cite{Minchin07}, which has high spatial resolution, and the data set obtained at Arecibo by  \cite{Haynes07}, which offers a better surface brightness sensitivity.  VirgoHI21 is the elongated condensation  at about  mid--distance in the HI tail. NGC~4254 is the spiral to the South. Its optical image from SDSS is superimposed. The inset shows the Position (Vertical axis) -- Velocity (Horizontal axis) diagram along the bridge.   {\it Right:}  The numerical model best reproducing the observations. The gaseous (in blue) and stellar (in white) components of the system are displayed at $t=750$~Myr. The corresponding PV diagram along the tidal tail is shown in the inset.  Another  PV diagram showing the influence of ``Object C" is also included. The scales are roughly the same for the observed system and the numerical model.
 \label{fig:model}}
\end{figure}

\begin{figure}
\includegraphics[width=\textwidth]{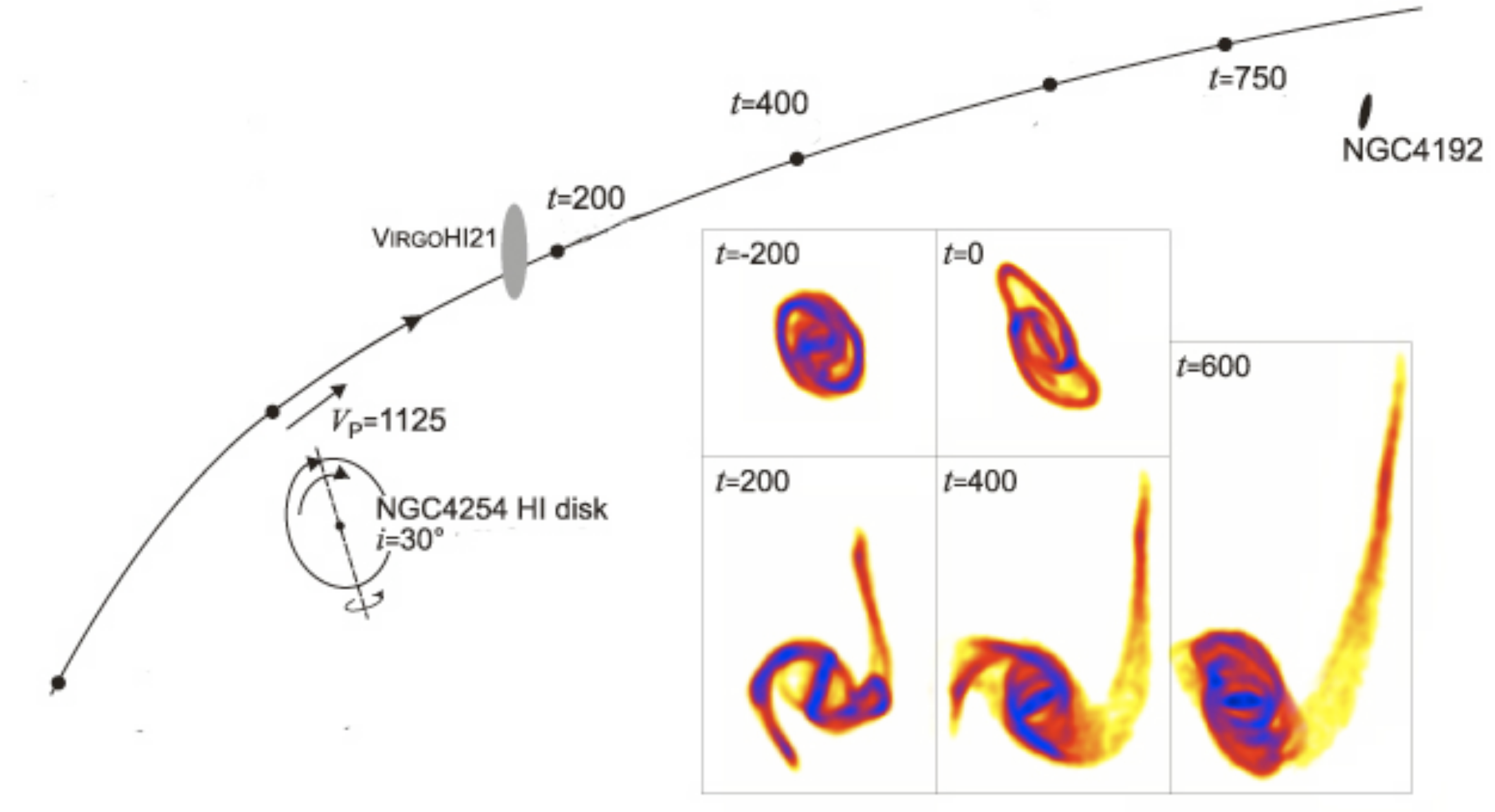}
\caption{The high-speed collision at the origin of VirgoHI21. The trajectory of the intruder in the plane of the sky  is shown together with different snapshots of the gas in the primary target, NGC~4254. Note that  in 3D, the intruder was behind NGC~4254 at pericenter and now approaches us.
\label{fig:trajectory}}
\end{figure}

VirgoHI21 is the prototype of Dark Galaxies \citep{Minchin05,Minchin07}, being one of the most massive dark objects thus far detected in HI. Its main characteristics are the following: an HI mass of $\sim 10^8$~M$_\odot$; the absence of any stellar counterpart even on deep HST images and a velocity gradient as large as 220~km~s$^{-1}$ (see Fig.~\ref{fig:model}) from which a dynamical mass of $\sim 10^{11}$~M$_\odot$ was  derived.  Whether such massive galaxies without any stellar counterpart may exist theoretically  has already been actively debated \citep{Taylor05,Davies06}. 
Mapping of the system in HI with the Westerbork Synthesis Radio Telescope revealed a bridge linking the putative Dark Galaxy to a spiral galaxy, NGC~4254, located at 150~kpc \citep{Minchin07}. Subsequent observations by \cite{Haynes07} with the Arecibo dish as part of the Alfalfa project indicate that it is in fact part of a larger HI filament. 

Our numerical model shown in  Figure~\ref{fig:model} is able to reproduce all of the above--mentioned features. It suggests that VirgoHI21 is simply tidal debris expelled from the parent galaxy, NGC~4254, during a fly--by  by a massive intruder. The collision occurred 750~Myr ago at a pericenter velocity of 1125~km~s$^{-1}$ (see Fig.~\ref{fig:trajectory}). The details of the model are presented in \cite{Duc07a}. Our high--velocity collision model addresses most of the issues that had been raised by \cite{Minchin07} against the tidal scenario, in particular:
\begin{itemize}
\item {\em There is no sign of an intruder galaxy.} According to our model, it  is already far away,  at a  projected distance of 400~kpc  to the NW of NGC~4254. 
A massive spiral is, in fact, present near this position, NGC~4192,  and could be a suitable candidate for the interloper although other  combinations of orbits/projection/age also reproducing the properties of VirgoHI21 could point out other candidates. 
\item {\em The system presents only a single--tail.}  Indeed,  a counter-tail has formed at the beginning of the collision, but by now it has had time to fall back.
\item  {\em The stellar disk of  NGC~4254 has only been weakly perturbed.} This is reproduced for the parent galaxy in high--velocity encounters.
\item {\em  The main tidal tail is almost entirely gaseous.} As observed, the only old stellar component in the model is located at the base of the tail, whereas the gas column density along the tail is too low to allow the formation {\em in situ}  of young stars.  
\item {\em   A condensation has formed within the tidal tail}. The orientation of the model was chosen so that its position matches that of VirgoHI21. 
\item {\em VirgoHI21 seems  kinematically independent}. In the model, this is an artefact: the sign of the velocity gradient changes as a result of projection effects and thus the  dynamical mass estimated from the observed velocity gradient is largely overestimated. 
\end{itemize}

Whereas the global kinematics of VirgoHI21 and its bridge/tail can be simply explained by streaming motions along a tidal structure,  when looked at in detail the model and observations show some minor differences which show up when comparing the  observed and simulated Position--Velocity diagrams: the velocity gradient towards VirgoHI21 is locally larger within the HI cloud, while in the model  it is not more enhanced at this location than further away near the tip of the tail. This local difference is actually not a real concern for the tidal  scenario as several additional processes may be put forward to  explain it: self-gravity in the HI condensation could contribute to enhance the velocity gradient, in addition to the streaming motions  (see Sec.~\ref{sec:tidal}); the velocity field of the tail may have been  perturbed by another  cluster member, in particular so-called ``Object C" (SDSS J121804.26+144510.4) an HI-rich star--forming dwarf just East of VirgoHI21. We have tentatively modeled its influence in our simulation and produced a Position--Velocity Diagram showing the  ``S--shape" profile which is characteristic of VirgoHI21 (see Fig~\ref{fig:model}).

Obviously, in a cluster of galaxies like Virgo, several other processes may have played a role,  such as ram--pressure \citep{Vollmer05} or the effect of the tidal field of the cluster itself \citep{Haynes07}. We claim however that the principle one, able to reproduce most of the properties of VirgoHI21, remains the  high--velocity collision. This explanation is much simpler than the Dark Galaxy scenario for which no convincing model yet exists that is able to reproduce the main characteristics of the system VirgoHI21/NGC~4254.

\section{Identifying a tidal origin for dark galaxy candidates}\label{sec:tidal}

Proving that VirgoHI21 is tidal debris (or, depending on the authors, a Dark Galaxy!)   is in principle an easy task. Indeed, the object is so extended that its internal kinematics may be studied in great detail. Besides, given its proximity,  deep HI surveys are able to reveal the faint HI filament linking it to its parent galaxy. This is not the case for the majority of Dark Galaxy candidates so far detected, which are compact and have much lower HI and (supposedly) dark masses. For those objects, building a realistic  numerical model is difficult. Nevertheless, if they are indeed of tidal origin, they should have some specific properties which we discuss in this section, based on the case study of the Tidal Dwarf Galaxy candidate, VCC~2062.

\subsection{A case study: VCC~2062}\label{sec:VCC2062}

\begin{figure}[ht!]
\includegraphics[width=11cm]{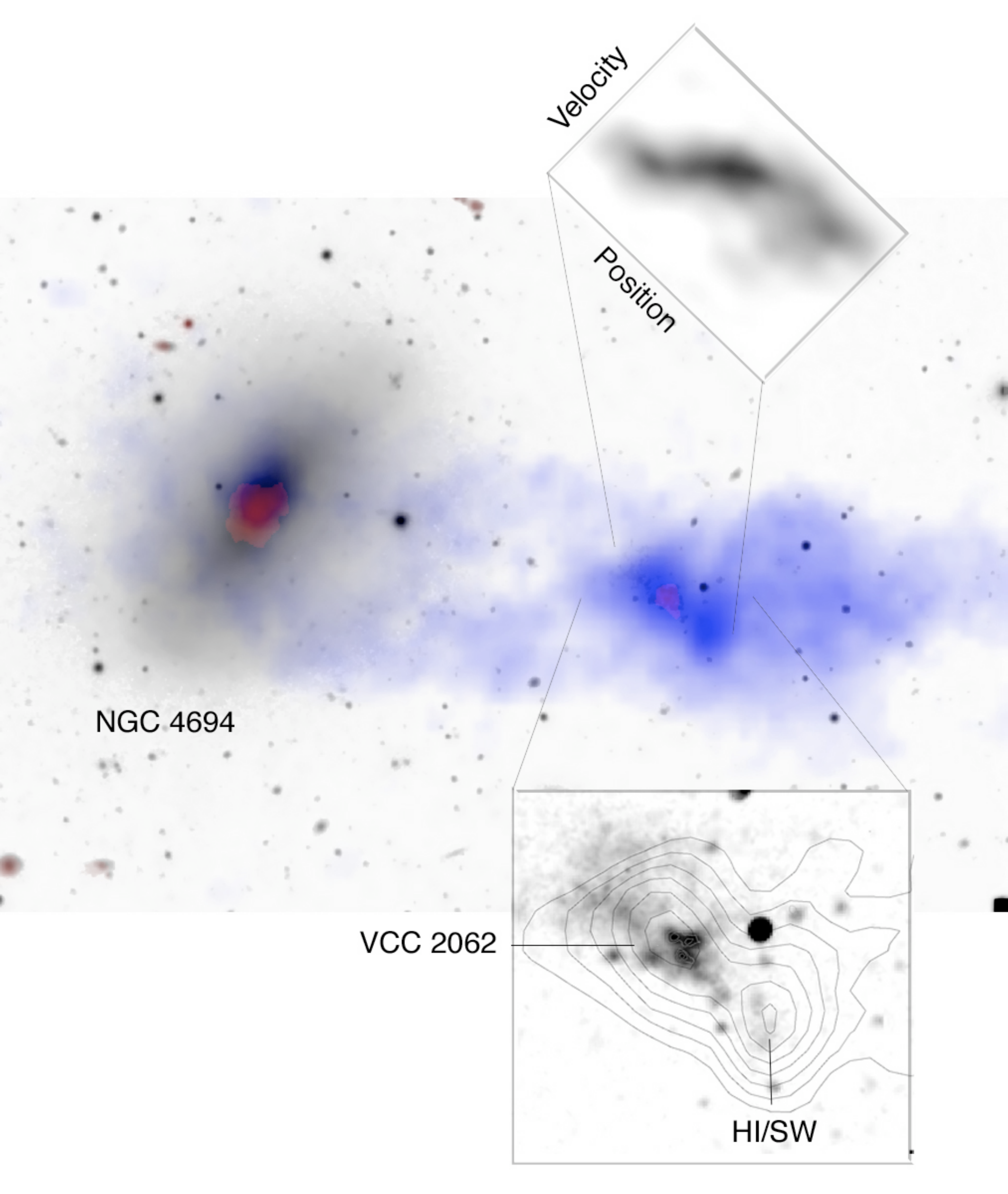}
\caption{The system VCC~2062/NGC~4694, located at a distance of about 1~Mpc from the core of the Virgo Cluster. The HI structure, as mapped by the VLA, is overlaid in blue on an SDSS image. The star--forming regions in the system, as traced by their GALEX far--ultraviolet emission,  are indicated in red. The lower panel is a zoom on the most prominent condensations in the HI tail. The HI contours (lowest: $4 \times 10^{20}$ cm$^{-2}$; highest: $10^{21}$ cm$^{-2}$) are superimposed on the optical image. VCC~2062 corresponds to the North-East(upper-left)  HI condensation. A Position-Velocity map of the HI structure is shown in the upper panel. The field of view is 9' $\times$ 6' (40~kpc  $\times$ 27~kpc)  \citep[adapted from][]{Duc07b}.
\label{fig:VCC2062}}
\end{figure}

Besides VirgoHI21, the Virgo Cluster hosts other cases of long HI filaments \citep[e.g.][]{Chung07}. One of the most massive ones, mapped long ago \citep{vanDriel89,Cayatte90}, is that emanating from the early--type galaxy NGC~4694, which we have studied in detail \citep{Duc07b}.  A  recent VLA HI map from the VIVA project \citep{Chung05} is shown in Figure~\ref{fig:VCC2062} together with an SDSS map of the region.

The tail, outside the optical body of  NGC~4694, contains about $10^9~$M$_{\odot}$ of HI. It is quite thick and overall has a complex morphology, showing at some places different branches and at others denser sub-structures. For most of its length, it has no optical counterpart. The exception is towards one of its most prominent condensations, where a low surface brightness stellar body is detected. It corresponds to the previously catalogued dwarf galaxy VCC~2062. There, the HI column density --- close to 10$^{21}$~cm$^{-2}$ --- is higher than the critical threshold for the on-set of star-formation. And, indeed, the dwarf exhibits star--forming regions as traced by H$\alpha$ and far--ultraviolet emission. Just to the South--West  of VCC~2062, another condensation has formed; its HI column density is even higher, but contrary to the previous object, it did not manage to recently  form stars and shows  at best an extremely faint optical counterpart. Why have these two nearby HI clouds such different properties? The answer comes from the kinematical analysis of the system. A Position--Velocity diagram based on the HI datacube reveals the presence of a well--defined velocity gradient (of about 45~km~s$^{-1}$ over 4~kpc) towards VCC~2062 while the SW condensation exhibits broad HI lines and no structured velocity field (see Fig.~\ref{fig:VCC2062}). In one case, the HI gas is kinematically decoupled from the rest of the tail, is gravitationally bound, rotating and forming stars like a normal galaxy, in the other it has not (yet) collapsed and has remained purely gaseous. 

\cite{Duc07b} discuss various hypotheses for the origin of the HI--structure around VCC~2062. According to the most attractive one, the HI in the tail would come from a small HI--rich galaxy that has been accreted by NGC~4694 a few hundred Myr ago. This latter galaxy exhibits several pieces of evidence for a past (minor) merger. In this scenario, the HI structure is a tidal tail and VCC~2062, which was formed in it, a Tidal Dwarf Galaxy \citep[see][for a review]{Duc07}. 

Actually, the main indication supporting the idea that VCC~2062 is a recycled object comes from follow--up optical and millimetre--wave spectroscopy. The emission lines of the HII regions of VCC~2062 indicate an oxygen abundance of  about solar (12+log(O/H)$=8.6-8.7$);  a strong signal from the CO(1--0) line has been  detected and mapped with the IRAM 30--m antenna. The measure of a high metallicity --- one dex higher than that inferred from the metallicity--luminosity relation --- and  the detection of molecular gas via   the CO emission is at odds with other star--forming classical dwarf galaxies in the Virgo Cluster, but is consistent with the tidal scenario.  These observations indicate that VCC~2062 is made out of material that has been pre--enriched in a larger galaxy. 
Finally, the ratio between the dynamical mass --- 3-4$\times 10^{8}~$M$_{\odot}$,  as inferred from the velocity gradient of the HI --- and the luminous mass --- consisting of HI, H$_{2}$ traced by CO and stars --- estimated to be  only of order 2--3, is much lower than that found for field, dark--matter dominated  dwarf galaxies. This excludes the hypothesis that  VCC~2062 is the stellar remnant of a pre--existing  more massive galaxy which would have been tidally disrupted (but would not have yet been accreted by NGC~4694). On the other hand, the fact that the dynamical and luminous masses are not equal reveals the existence of some missing matter (see Sec.~\ref{sec:dark}).

\subsection{Criteria for a tidal origin}\label{sec:preenriched}

VCC~2062 is an example, among others,  of tidal debris that was massive enough to become self--gravitating and form a new generation of galaxies. Tidal Dwarf Galaxies are rather rare objects in the nearby Universe, given the many criteria that are required to form them \citep{Bournaud06}; however they are particularly useful laboratories to investigate a number of key questions, among them how to identify tidal debris once it has travelled far from their parent galaxies. Here is a list of such criteria:

\begin{itemize}
\item {\em The presence  of pre-enriched gas.} Whereas genuine Dark Galaxies should be made of almost pristine gas (since no generation of stars has polluted their ISM), tidal debris in general and TDGs in particular have inherited from their parent galaxy their metal content. Contrary to common belief, tidal material may partly come from regions initially located inside the optical radius, and have thus a metallicity higher than that of the outer disk of spirals. The typical metallicity measured in TDGs is about half solar instead of one tenth solar in the extreme outer regions of  spirals \citep{Ferguson98}.
 The metallicity of HI gas is unfortunately not easy to measure. If  the HI is dense enough to collapse and form stars,  element abundances can be determined from optical lines emitted by  the surrounding ionized  gas. 
 This is the case in TDGs or fainter so--called intergalactic HII regions. If a background source such as a quasar  is by chance present behind the gas  cloud, the metallicity can be directly determined via absorption in the HI component. 
 \item {\em The presence of molecular gas} together with the atomic hydrogen. Tidal collisions not only strip the HI, but all the ISM of a galaxy, in particular its molecular gas. Moreover, H$_{2}$ may also form locally in HI condensations, as shown by \cite{Braine00}. This may occur even before stars are formed. For instance, CO was detected in the dense, star--less, HI cloud SW of VCC~2062. 
 \item {\em The presence of dust}, either in emission, e.g. PAHs,  in star--forming regions \citep{Higdon06,Boquien07}, or in absorption through the reddening of the background galaxies \citep[e.g.][]{Xilouris06}.
 \item {\em A dynamical mass of the same order as (but not necessarily equal to)  the luminous mass}. As discussed in the next section, cosmological dark matter, presumably located in a pressure--supported halo around galaxies, is present in only small proportion in tidal tails. Tidal debris is not as strongly dark matter dominated as classical galaxies and Dark Galaxies. Such a property can be checked if their dynamical mass can be measured and compared to the luminous one. This actually requires the HI cloud to be gravitationally bound  and contamination by streaming motions to have been subtracted, as  emphasized in the section on VirgoHI21. 
 \end{itemize}
 
 In practice, the above mentioned  criteria apply well for the most massive tidal debris, in particular for TDGs, but may be more difficult to check for the faintest ones that have no optical counterpart, e.g.  the high-velocity clouds in the Local Group.   The case study of VCC~2062 shows that whenever a cloud becomes kinematically independent and starts rotating, it forms stars and becomes visible in the optical. The probability to be able to measure the metallicity of a quiescent, star-less HI cloud, thanks to  a background quasar, is very low. The detection of molecular gas in tidal debris via the millimetre--wave CO line, although facilitated by the relatively high metallicity remains, however very challenging and requires the use of a big antenna.

\section{A dark component in tidal debris}\label{sec:dark}
The idea of collisional debris can be put forward to argue against the existence of  ``Dark Galaxies". In addition, it can tell a lot about  the distribution and nature of Dark Matter in and around galaxies. First of all, \cite{Bournaud03} argued that the most massive tidal condensations --- the progenitors of TDGs --- could only be formed if the parent galaxies had an extended dark matter halo around them. Otherwise, the potential well of the system and   tidal forces are such that the ISM gets  diluted  along the tails, allowing the formation and growing of  only small condensations  \citep{Duc04b}. 
Second, as argued before, collisional debris by itself doesn't accrete dark matter from cosmological {\em haloes}. Thus, if they contain a certain amount of missing matter, it should also be present in the rotating disk of the progenitor where all tidal material comes from. Checking this is particularly difficult, though; we could achieve it so far in only two systems for which we had the required data to determine with high enough  precision  the total (dynamical) mass and compare it to the luminous mass: VCC~2062 which was presented earlier and NGC~5291, studied in detail by \cite{Bournaud07} and briefly presented below. 

\subsection{The giant collisional ring around NGC~5291}
NGC~5291 is an early--type galaxy located at the edge of a cluster of galaxies. It is surrounded by a prominent HI ring with a diameter exceeding 160 kpc and a mass as high as $5 \times 10^{10}~$M$_{\odot}$. Observations and the numerical model of  \cite{Bournaud07} favor a collisional origin: the ring was formed following the off-center direct hit at a velocity of 1250 km~s$^{-1}$ of the gaseous disk of NGC~5291 by a massive intruder, 360 Myr ago. 
At several locations along the ring, the gas has collapsed and formed stars. About 30 of such star--forming regions have been identified thanks to their ultraviolet, H$\alpha$ and mid--infrared emission \citep{Boquien07}. Molecular gas was detected in two of them \citep{Braine01}. The internal kinematics of the three most prominent condensations could be studied  using Fabry-Perot H$\alpha$ and high-resolution VLA B--array HI datacubes. They exhibit flat rotation curves from which a dynamical mass has been derived. For the most massive object, it amounts to $30\pm8.6  \times 10^{8}~$M$_{\odot}$. The error of the method has been checked using numerical simulations. It turns out that all three objects have a total mass three times higher than the luminous one as inferred from 21~cm (HI), CO (tracing the H$_{2}$)  and optical/near-infrared (tracing the stars) observations. This would indicate that, contrary to expectations, a missing mass component is present in this collisional debris, a result also suggested by the study of the TDG VCC~2062. According to our  most conventional scenario,  the dark component is made of invisible baryons, perhaps cold molecular gas \citep{Pfenniger94} not accounted for by classical CO observations. Alternatively, the way the masses are determined from the rotation curves is wrong, as suggested recently by \cite{Milgrom07} and \cite{Gentile07}, who present models consistent with the theory of Modified Newtonian Dynamics (MOND).

\subsection{Dark baryons in spiral disks}
If dark baryons are found in collisional debris, they should also be present in the disk of their parent galaxies. This would suggest that not all of the missing baryons of the local Universe (see the review  of 
C. Impey in this volume) should be looked for in the Warm--Hot (WHIM) gas around galaxies. The content of the ISM might have been underestimated, as also claimed by \cite{Grenier05} for the solar neighborhood.
We are currently implementing this possible additional component in our numerical model to investigate its impact on the stability of spiral disks and determine its behaviour,  would a collision occur.

\begin{acknowledgments}
Several people have contributed to the various projects presented here, in particular J. Braine, U. Lisenfeld and  M. Boquien. They are warmly thanked. We also appreciated the opportunity offered by the organizers of this IAU symposium, in particular J. Davies and R. Minchin,  to present and discuss our results in a stimulating and  friendly atmosphere. 
\end{acknowledgments}



\end{document}